\documentclass{article}
\usepackage{spconf,amsmath,graphicx,hyperref}

\title{Shared Multi-modal Embedding Space for Face-Voice Association}
\name{Christopher Simic, Korbinian Riedhammer, Tobias Bocklet}
\address{Technische Hochschule Nuernberg, Nuremberg, Germany}
\usepackage{multirow}
\usepackage{booktabs}    
\usepackage{amssymb}
\usepackage{multirow}
\begin{document}
\ninept
\renewcommand{\baselinestretch}{0.87}\normalsize

\maketitle
\begin{abstract}

The FAME 2026 challenge comprises two demanding tasks: training face–voice associations combined with a multilingual setting that includes testing on languages on which the model was not trained. 
Our approach consists of separate uni-modal processing pipelines with general face and voice feature extraction, complemented by additional age-gender feature extraction to support prediction. 
The resulting single-modal features are projected into a shared embedding space and trained on an Adaptive Angular Margin (AAM) loss. Our approach achieved first place in the FAME 2026 challenge, with an average Equal-Error Rate (EER) of 23.99\%.

\end{abstract}

\section{Introduction}


Speaker recognition and identification tasks are well studied and latest approaches achieve remarkable results close to 100\% recognition rate. Typically, a strict distinction is made between audio-based~\cite{Desplanques2020ECAPATDNNEC, spk_rec_2011, snyder2018xvector, dehak2011ivector} and image-based~\cite{schroff2015facenet,parkhi2015deep,Cao2017VGGFace2AD} approaches for speaker verification. Some, few approaches combine audio and visual information jointly to improve recognition performance~\cite{simic2025combining,tao2023deepcleansing}.

The FAME 2026 Challenge~\cite{moscati_fame2026} addresses an even more challenging problem: using combinations of voice samples and facial images, the model is supposed to determine whether they belong to the same speaker or to different speakers.
To evaluate model performance, the Equal Error Rate (EER) es reported, defined as the operating point at which the false-positive and false-negative rates are equal.

While earlier approaches to face–voice association rely on joint processing of both modalities~\cite{Fame_baseline, tao2024multi}, we present an approach that processes audio and image data separately. 
The modality separated embeddings are then mapped into a shared embedding space, aiming to minimize the angular distances between samples from same speakers, while maximizing the distance between different speakers.

\section{Data}

We use three complementary datasets: VoxCeleb2 \cite{data_vox2}, CommonVoice \cite{data_commonvoice}, and the official challenge dataset MavCeleb.

The MavCeleb V3 set contains 3,812 English and German utterances from 58 speakers. Following the challenge protocol, 50 speakers are used for fine-tuning and development, while the remaining 8 are held out for final evaluation. Due to the limited dataset size, we employ a seven-fold cross-validation scheme. For evaluation, hyperparameter tuning, and early stopping, we construct image–voice verification trials for each fold. In this paper, we denote this set, generated from the training samples, as the \textit{dev set}.

VoxCeleb2 provides over 1 M utterances from 5,994 speakers in multiple languages, with a majority of English samples.
For visual processing we extract every 25th video frame.
To comply with the heard/unheard challenge settings, we prepare three training variants: (I) a full set, (II) a German-excluded set (german-unheard) and (III) an English-excluded set (english-unheard).

CommonVoice, a large crowd-sourced speech corpus with broad linguistic and demographic coverage, supplies age and gender labels for most recordings. We use the English, German, Danish, Dutch, Spanish, French, and Italian subsets to train dedicated audio-based age–gender prediction models, which also follow the heard/unheard challenge rules.

\section{Proposed Approach}

\begin{figure}
  \centering
  \includegraphics[width=0.7\linewidth]{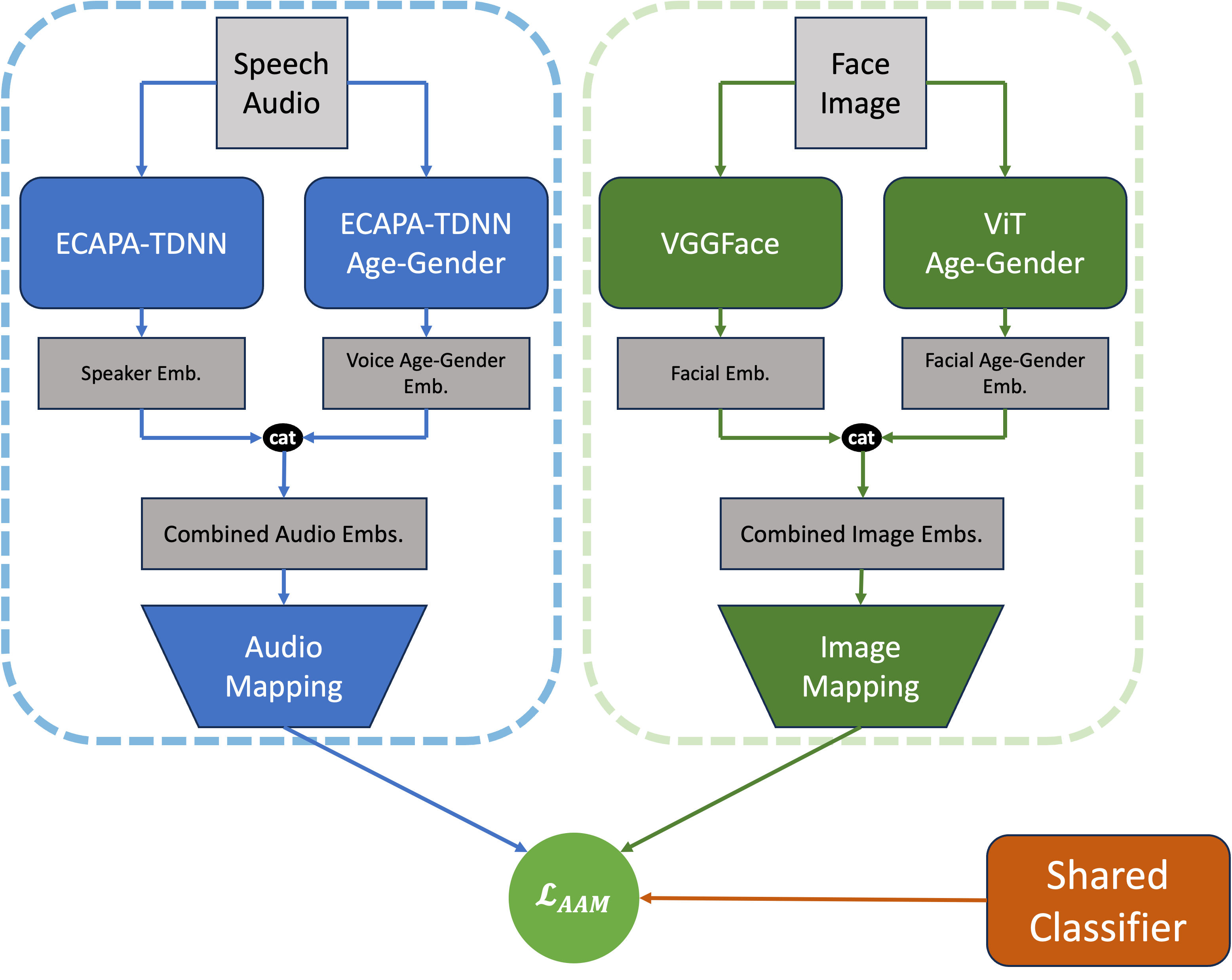}
  \caption{
  Overall model architecture, comprising separate modality-specific processing pipelines, dedicated mapping layers and a shared classifier. Both pipelines are optimized using the AAM loss.}
  \label{fig:model_overview}
\end{figure}

For our approach we rely on a separate processing of both modalities (audio and image), as illustrated in Figure~\ref{fig:model_overview}

\subsection{Audio Processing}

Our audio processing pipeline (blue frame) consists of two modules: A standard ECAPA-TDNN~\cite{Desplanques2020ECAPATDNNEC} that is trained to separate different speakers and a small ECAPA-TDNN, that is trained on CommonVoice to predict age and gender for given audio samples.

For our approach, we use three different ECAPA-TDNN speaker embedding models, according to the challenge rules. The standard ECAPA-TDNN model from Speechbrain suits the heard scenarios. For the unheard scenarios we prepare ECAPA-TDNN models that are trained on the reduced VoxCeleb2 dataset after removing samples from each prohibited language (English/German). These models consist of 20.8M parameters. The embeddings for subsequent processing are taken from the layer before the final output layer with a dimension of 6144.

Our custom voice age-gender model is based on the same architecture, but with reduced channel dimensions by factor 4. For age and gender prediction we add two heads, each consisting of a two-layer MLP (Multi-Layer-Perceptron). This leads to a model sizes of 6.5~M parameters. 
Again we prepare three different models: One is trained on all prepared CommonVoice languages, which may be used for the heard scenarios. Two different models are trained after removing English respectively German samples to match the challenge unheard rules. The embeddings for further processing are taken from the last shared layer with a dimension of 1536.

As illustrated in Figure~\ref{fig:model_overview} both embeddings are concatenated and subsequently processed through a single linear mapping-layer to reduce the dimension to 192.

\subsection{Image Processing}

For image processing, our approach follows a similar pipeline as for audio processing, with two separate modules (green frame).

For general face feature extraction we adopt the standard VGGFace~\cite{parkhi2015deep} model. The embeddings are taken from the layer before the final layer, with an embedding dimension of 4096. 

Similar to the audio pipeline, we add a second module to extract information about age and gender. Therefore, we adopt a Vision-Transformer (ViT) based model~\cite{age_gender_prediction_2025}, that achieves a gender accuracy of 94.3\% and an average age error of 4.5 years. The embeddings are taken from the last shared layer with a dimension of 768.

Also for image processing both embeddings are concatenated and processed through a single linear mapping-layer with an output dimension of 192.

\subsection{Training Strategy}

For face-voice association all ECAPA-TDNN, VGGFace und ViT modules remain frozen, while only the mapping-layers are trained. It turned out to be beneficial to mask large parts of the input embedding by adding a dropout layer with a dropout probability of 0.9, to prevent overfitting. We assume that this is related to the high embedding dimensions that may contain redundant information. 
To project both modalities into a shared embedding space, both are trained jointly using a shared classifier (Figure \ref{fig:model_overview}, orange block) optimized with an Adaptive Angular Margin (AAM) loss. This loss encourages the model to produce embeddings with small distances for same-speaker pairs and near-orthogonal embeddings for different speakers. For score computation, our system uses cosine similarity.

\subsection{Jointly Attention-based Processing}


In addition to the modality-separated pipelines, we investigate a unified fusion strategy. After independent pre-processing of audio and image inputs, the resulting embeddings are segmented and reshaped into sequence representations. These audio and image sequences are then passed through a two-layer cross-attention architecture, which fuses information across both modalities.
Finally, a linear projection to a single output neuron determines whether a given image–audio pair originates from same or from different speakers.

\section{Results}

In this section, we present and discuss several training strategies and architectural variants that we evaluated on the final challenge test set.


\subsection{Combined Datasets}

We start with combining the samples from VoxCeleb2 and MavCeleb for a joint training on the characteristics of both datasets. For hyperparameter tuning and early stopping we prepared a 5\% split from VoxCeleb2 to prevent overfitting. This training strategy led to a submission result of 24.57\% EER.

\subsection{Pre-Training vs. Fine-Tuning}

A more sophisticated training strategy relies on pre-training on VoxCeleb2 with subsequent fine-tuning on MavCeleb. 
The fine-tuning is performed in a cross-validation setup on the prepared seven MavCeleb training folds.
Evaluation on the \textit{dev set} samples, generated from MavCeleb training data, reveals an average EER of 24.77\% after pre-training and small deterioration after subsequent fine-tuning with an average EER of 24.98\%.
The same behavior can be observed for the challenge test submission, with average EER after pre-training of 24.93\% and 25.30\% after fine-tuning.

\subsection{Full Dataset for Heard Scenario}
\label{subsec:full_ds_heard}

For both heard scenarios we investigate the influence of including all VoxCeleb2 data during training, since not restrictions are made for these scenarios.
The evaluation for english-heard on our \textit{dev set} reveals an EER of 24.07\%, while the model, trained without German samples achieves poorer results with EER 24.94\%. We assume that additional German samples provide more variability, leading to better results.
For german-heard the evaluation of a model, trained on all data reveals poorer performance with an EER of 25.25\%, compared to 24.60\% for a model, that was trained without English samples.
Since the number of English samples dominates the VoxCeleb2 dataset, the training may bias the model towards english, resulting in reduced performance on German speech.
Therefore we submitted a combination, with the all data model for english-heard and the model that was trained without English samples for german-heard.
For english-unheard the model without any English samples during training is taken and for german unheard the no German sample model. This combination leads to improvements compared to previous strategies with an average EER of 24.36\%.

\subsection{Fine-tuned Model for Unheard Scenarios}

Although previous experiments demonstrate, that fine-tuning does not improve performance in heard scenarios, we decided to evaluate MavCeleb fine-tuned models for the unheard conditions, while keeping model selection for heard scenarios as described in Section~\ref{subsec:full_ds_heard}.
This combination yields consistent gains and reduces the final challenge EER to 23.99\% (for detailed results, see Table~\ref{tab:main_results}). 
This trend is also observed on our \textit{dev set}.

We attribute this behavior to two key aspects of the target domain.
First, large-scale pre-training on datasets that already include the target language (e.g., VoxCeleb2) equips the model with broad variability and strong generalization. Subsequent fine-tuning on already heard languages may lead to overfitting, which is the reason why the models for both heard scenarios perform better immediately after pre-training.


For the unheard scenarios, the domain gap between pre-training data and final test data is larger because the target language is not present in the VoxCeleb2 data, and the recording conditions and face crop may differ between VoxCeleb2 and MavCeleb test data. Fine-tuning therefore reduces the domain gab by aligning the models closer to the recording conditions and face crop, while still maintaining the strict policy for unheard languages.

\begin{table}[h]
\centering
\begin{tabular}{lccc}
\toprule
 & Eng.\ test & Ger.\ test & Overall \\ 
\midrule
English\ train & 30.6* & 17.4 & \multirow{2}{*}{23.99} \\
German train  & 30.1 & 17.9 &      \\
\bottomrule
\end{tabular}
\caption{Evaluation results for English and German. * The results for english-heard (English train / English test) are predicted by the model, that was trained on all VoxCeleb2 data.}
\label{tab:main_results}
\end{table}


\subsection{Jointly Attention-based Processing}

Our alternative architecture, which processes and fuses both modalities jointly through a multi-layer cross-attention module, performed poorer than our modality-separate approach. This model achieved an average EER of 28.92\% after submission, which would still have ranked second in the challenge.

\bibliographystyle{IEEEbib}
\bibliography{strings,refs}

\end{document}